# Improving polyhydroxyalkanoates production in phototrophic mixed cultures by optimizing accumulator reactor operating conditions

J.C. Fradinho[*], A. Oehmen[†], M.A.M. Reis

UCIBIO-REQUIMTE, Department of Chemistry, Faculty of Sciences and Technology, Universidade NOVA de Lisboa, 2829-516 Caparica, Portugal

**Abstract**

Polyhydroxyalkanoates (PHAs) production with phototrophic mixed cultures (PMCs) has been recently proposed. These cultures can be selected under the permanent presence of carbon and the PHA production can be enhanced in subsequent accumulation steps. To optimize the PHA production in accumulator reactors, this work evaluated the impact of 1) initial acetate concentration, 2) light intensity, 3) removal of residual nitrogen on the culture performance. Results indicate that low acetate concentration (<30CmM) and specific light intensities around 20W/g$X$ are optimal operating conditions that lead to high polyhydroxybutyrate (PHB) storage yields (0.83±0.07 Cmol-PHB/Cmol-Acet) and specific PHB production rates of 2.21±0.07 Cmol-PHB/Cmol $X$ d. This rate is three times higher than previously registered in non-optimized accumulation tests and enabled a PHA content increase from 15 to 30% in less than 4h. Also, it was shown for the first time, the capability of a PMC to use a real waste, fermented cheese whey, to produce PHA with a hydroxyvalerate (HV) content of 12%. These results confirm that fermented wastes can be used as substrates for PHA production with PMCs and that the energy levels in sunlight that lead to specific light intensities from 10 to 20W/g$X$ are sufficient to drive phototrophic PHA production processes.

[*]Corresponding author
E-mail address: j.fradinho@campus.fct.unl.pt (J.C. Fradinho).
[†]Present address: School of Chemical Engineering, University of Queensland, Brisbane, QLD, 4072, Australia

**Introduction**

Polyhydroxyalkanoates (PHAs) are intracellular biopolymers synthesized and accumulated by several microorganisms as carbon and energy reserves. These biopolymers have attracted significant interest from the industry and research community because they present physicochemical properties similar to conventional plastics [1,2]. Moreover, PHA can answer society's request for an environmentally sustainable bioplastic production process since PHA is completely biodegradable [3] and can be produced from agro-industrial residues [4]. Presently, the PHA available in the market is produced in pure culture systems that require intensive aeration, media/equipment sterilization and strict control of reactor operation. This leads to high production costs, in comparison to the synthetic plastic production, which limits the wider commercialization of microbial produced PHA [5].

In order to decrease PHA production costs and bring this sustainable biopolymer into the market, the utilization of mixed microbial cultures (MMCs) that can use cheap complex residues as feedstock in open conditions has been proposed [6]. Typically, these MMCs are capable of producing PHA through the application of transient carbon availability conditions - so-called Feast and Famine (FF) strategy - which continuously selects for organisms capable of storing PHA during a short feast phase and consuming it during a long famine phase. Over time, this selection pressure enriches the culture in organisms with high PHA storage capacity [7]. This FF strategy has been applied to a multitude of different feedstocks like agro-industrial wastes (e.g. fruit pomaces, animal litter, cheese whey, glycerol) or food and urban wastes (e.g. used cooking oil), where very successful results were obtained in terms of both productivity and accumulation capacity (as reviewed in [4, 5, 8]). However, PHA production with MMCs has been mainly restricted to the utilization of aerobic organisms, while in fact, the diversity of bacterial species that can produce and accumulate PHA is much wider.

In recent years, studies with PHA producing phototrophic systems have proposed the utilization of phototrophic mixed cultures (PMCs) as means of decreasing operational costs [9]. Phototrophic organisms can draw energy from sunlight and by not requiring oxygen to produce ATP, aeration is nonessential, and the high costs associated with system's aeration can be eliminated. Initial studies with PMCs also applied the FF strategy to obtain PHA storing phototrophic bacteria, and by using acetate as carbon source, PHA accumulation values of 20% and 30% g PHA/g volatile suspended solids (VSS) were obtained under continuous illumination [9] and transient dark/light conditions [10], respectively. Nevertheless, a recent study proposed a new selection strategy for PMCs, where the culture is maintained in the continuous presence of carbon [11]. This permanent carbon feast strategy is based on the specific characteristics of anoxygenic phototrophic bacteria that do not release oxygen during phototrophy but can phototrophically produce ATP to take up external carbon. Because the system is operated in anaerobic conditions, the organisms must activate internal mechanisms to oxidize reduced molecules produced during cell metabolism (like NADH, NADPH). One of these mechanisms is through the accumulation of PHA that requires its precursors' reduction during the polymer formation. Therefore, the permanent feast strategy selects organisms that are capable of regulating internal reducing power *via* PHA production. The results obtained during the development of this permanent feast strategy indicated that permanent feast regimes enable the selection of PMCs which have high PHA accumulation capacity when PMCs are exposed to higher light availability. Indeed, [11] reported PHA contents of 60% PHA/VSS in accumulation tests conducted with high light availability, while in the selector reactor, operated at low light availability, the PHA content averaged 3-5%. Therefore, an operational strategy was proposed, consisting in operating the selector reactor with low light availability (low light SBR) and conducting the PHA accumulation in separate reactors under light optimized conditions (high light accumulator). The present work intends precisely to evaluate this operating strategy and test different operating conditions in the accumulator reactors, like illumination, carbon concentration and nitrogen availability, in order to maximize PHA production by PMCs. Also, in view of the fact that, so far, PMCs have only been tested with synthetic feed, this work evaluated for the first time the possibility of using a real waste, fermented cheese whey (FCW), as feedstock for PHA production in the accumulator reactor.

## 2. Materials and methods

### 2.1 PMC selector reactor operation

The PMC studied in this work was obtained from the phototrophic selector reactor operated in [11] under a carbon permanent feast regime, using acetate as carbon source. The PMC was subjected to the same operating conditions with the exception of the reactor vessel that in the present study had a working volume of 4.4L and was internally illuminated by a halogen lamp (200 W) at a light intensity of 150 W/m$^2$, which corresponded to a volumetric light intensity of 1.3 W/L of culture broth (the reactor in [11] was externally illuminated with a volumetric light intensity of 1.8 W/L).

The PMC operation was performed under continuous illumination in a SBR (henceforth called selector SBR) with 24h cycles. At each cycle, the selector SBR was fed with equal amounts of culture medium (733 mL containing per L: 0.8 g MgSO$_4$.7H$_2$O, 1.6 g NaCl, 2.2 g NH$_4$Cl, 0.2 g CaCl$_2$.2H$_2$O, 16.4 g NaAcetate.3H$_2$O, 20 mL iron citrate solution (1.0g/L), 4 mL trace element solution) and phosphate medium (733 mL containing per L: 0.13 g KH$_2$PO$_4$ and 0.17 g K$_2$HPO$_4$), which corresponded to an organic loading rate of 1.3 g COD/L d and a C:N:P molar ratio of 124:20:1. At the end of each cycle, 1466 mL of the continuously stirred PMC were wasted, resulting in a hydraulic retention time (HRT) and sludge retention time (SRT) of 3 days. Temperature was controlled at 30ºC and argon was continuously sparged (10 mL/min) to prevent surface aeration. pH was controlled at 6.5 using 0.5 M HCl.

**2.2 Optimization of accumulation reactor operation**

In order to determine the best conditions to operate the accumulator reactor, three parameters were sequentially tested: 1) initial acetate concentration, 2) light intensity, 3) removal of residual nitrogen.

**2.2.1 Effect of initial substrate concentration**

To evaluate the effect of the initial carbon concentration on the culture performance during the accumulation step, two acetate concentrations were tested. These concentrations were: A) the concentration typically present in the SBR at the beginning of the reactor cycle after feeding (higher concentration) and B) the concentration left by the residual acetate present in the SBR at the end of the cycle (lower concentration). To conduct these tests, mixed liquor withdrawal (300 mL) was collected from the selector reactor at the end of the cycle and placed in a separate accumulator reactor (working volume 450 mL). In the test with higher acetate concentration (test A), the mixed liquor withdrawal was fed with 150 mL of fresh medium (75 mL of culture medium + 75 mL of phosphate medium). The same occurred for the test with lower acetate concentration (test B) with the exception that the culture medium lacked acetate. The accumulator reactor was illuminated with a volumetric light intensity of $1.1 \pm 0.02$ W/L and operated under the same controlled temperature, pH and argon conditions as the selector reactor. For each concentration, a duplicate test was conducted for a period of 5 hours.

**2.2.2 Effect of light availability**

In order to evaluate the effect of light availability on the culture performance, the accumulator reactor was operated under the same conditions of the test B reactor in section 2.2.1 (low initial acetate concentration) and four increasing volumetric light intensities were tested: $1.1 \pm 0.0$ W/L, $4.7 \pm 0.0$ W/L, $10.6 \pm 0.6$ W/L, $20.7 \pm 0.7$ W/L. For each light intensity, a duplicate test was conducted for a period of 5 hours.

**2.2.3 Effect of residual nitrogen removal**

To evaluate the effect of conducting PHA accumulation assays in the absence of nitrogen, 333 mL of mixed liquor withdrawal collected at the end of the SBR cycle was centrifuged (9000g, 4 min, 20ºC) and the supernatant was removed. The obtained centrifuged biomass was re-suspended in 500 mL of fresh medium (250 mL culture medium lacking acetate and ammonia + 250 mL of phosphate medium). Periodic pulses of a concentrated solution of acetate (1.32 Cmol/L) were given along the test with each pulse leading to initial acetate concentrations in the reactor lower than 3.5 CmM but maintaining the reactor under the permanent presence of carbon. The accumulator reactor was illuminated with a volumetric light intensity of 20.0 W/L and operated under the same controlled temperature, pH and argon conditions as the selector reactor.

For performance comparison, an extra accumulation test was conducted using 500 mL of mixed liquor withdrawal that was directly placed in the accumulator reactor with no addition of fresh medium. To further evaluate the effect of carbon pulse addition, periodic pulses of an acetate solution (1.32 Cmol/L) were added, as mentioned above, after the residual acetate became depleted. This accumulator reactor was illuminated and operated as previously described for the assay with nitrogen removal.

**2.3 PHA accumulation with fermented cheese whey as substrate**

To evaluate the culture capability of using fermented cheese whey (FCW) as substrate for PHA production, two accumulation tests were conducted in the same conditions as in section 2.2.3 (with supernatant removal) but with the pulse addition of synthetic FCW or FCW. For each test, six pulses of synthetic or FCW were given in a way that the initial acetate concentration would be ≈ 3.0 CmM. Table 1 indicates the composition of the synthetic medium and FCW used in the tests. The FCW used in the present work came from a continuous stirred tank reactor (CSTR) operated in anaerobic conditions, HRT of 1 day, pH 6 and fed with cheese whey that was mostly composed of lactose (78.4%, w/w), proteins (13.6%, w/w) and fats (1.2%, w/w) with an organic loading rate of 15g COD $L^{-1}$ $d^{-1}$. (For further details on the CSTR operation please see [12]).

Table 1 – Composition of the synthetic fermented cheese whey (FCW) and FCW used in the accumulation tests. The synthetic FCW only simulates the organic acid and ethanol content of the FCW.

|  | Synthetic FCW | | FCW | |
| --- | --- | --- | --- | --- |
|  | Concentration (Cmmol/L) | % C | Concentration (Cmmol/L) | % C |
| Lactic Acid | 56 | 17 | 52 | 16 |
| Acetic Acid | 132 | 41 | 146 | 45 |
| Propionic Acid | 43 | 13 | 45 | 14 |
| Butyric Acid | 68 | 21 | 74 | 23 |
| Valeric Acid | 14 | 4 | 19 | 6 |
| Ethanol | 9 | 3 | 16 | 5 |

**2.4 Analytical methods**

PHA determination was performed by gas chromatography using the method described in [9]. Organic acids, ethanol, lactose and phosphate concentrations were determined by high-performance liquid chromatography (HPLC) using an RI detector and an Agilent Metacarb 87H column. 0.01 N sulfuric acid was used as eluent with an elution rate of 0.6 mL/min and a 30ºC operating temperature.

Total carbohydrates hydrolysable to glucose (i.e. bacterial glycogen and algae starch) were determined using the method described by [13], with minor modifications described in [9].

Ammonia was determined by colorimetric methods implemented in a flow segmented analyser (Skalar 5100, Skalar Analytical, The Netherlands). Volatile suspended solids (VSS) were determined according to Standard Methods [14]. The light intensity provided during the tests was measured using a Li-COR light meter LI-250 A equipped with a pyranometer sensor LI-200 SA.

Pigments extraction was performed by centrifuging 8 mL of mixed liquor (8000g, 5 min), and adding 8 mL of 100% ethanol to the biomass pellet. The biomass was then vortexed and incubated overnight at room temperature and in dark conditions. Afterwards, samples were again centrifuged at 8000g for 5 minutes and the absorbance spectrum of the supernatant was measured from 290 nm to 900 nm using an Ultrospec 2100 pro Amersham spectrometer.

**2.5 Calculation of kinetic and stoichiometric parameters**

The biomass PHA content was calculated as a percentage of VSS on a mass basis (%PHA = 100 × g PHA/ g VSS), where VSS includes active biomass ($X$), PHA and total carbohydrates. Active biomass was calculated by subtracting PHA and total carbohydrates from VSS. To determine the active biomass concentration in Cmol, it was assumed that the generic formula of biomass $CH_{1.8}O_{0.5}N_{0.2}P_{0.02}$ can be applied to PMCs.

The maximum specific substrate uptake rate (-$q_S$ in Cmol Acet/Cmol $X$ d), maximum specific PHA production rate ($q_{PHA}$ in Cmol PHA/Cmol $X$ d), maximum specific carbohydrates production rate ($q_{Carbs}$ in Cmol Carbs/Cmol $X$ d), maximum specific phosphate consumption rate (-$q_{PO4}$ in Pmol $PO_4$/ Cmol $X$ d) and maximum specific ammonia consumption rate (in Nmol $NH_4$/ Cmol $X$ d) were determined by adjusting a linear regression line to the experimental concentrations determined along the cycle and dividing the slope by the concentration of active biomass at the beginning of the cycle ($X_i$).

The yield of PHA per substrate consumed ($Y_{PHA/S}$ in Cmol PHA/ Cmol Acet) was calculated by dividing the amount of PHA formed by the amount of acetate consumed. The yield of phosphate per biomass ($Y_{P/X}$ in Pmol $PO_4$/Cmol $X$) was calculated by dividing the amount of phosphate consumed by the amount of active biomass grown in the same period.

The specific light intensity of the culture (W/g $X$), was calculated by dividing the volumetric light intensity of the culture broth (W/L) by the active biomass concentration (g $X$/L).

## 3 Results and discussion

### 3.1 Phototrophic mixed culture operation

In the present work, the PMC selected in [11] under the permanent feast strategy was transferred to a new selector reactor and operated under conditions that kept the culture with kinetic and stoichiometric parameters similar to the ones obtained in the previous work [11] (Table 2).

Table 2 - Kinetic and stoichiometric parameters of the PMCs performances in the selector reactor of the present study and in [11].

|  | W/L | $q_{PHB}$ | $-q_S$ | $-q_{PO4}$ | $Y_{PHB/S}$ | $Y_{X/S}$ |
|---|---|---|---|---|---|---|
| Selector SBR[a] (present work) | 1.3 | 0.02 (0.02) | 0.78 (0.07) | 0.029 (0.002) | 0.03 (0.02) | 0.53 (0.04) |
| Selector SBR[b] [11] | 1.8 | 0.05 (0.04) | 0.69 (0.08) | 0.014 (0.002) | 0.07 (0.05) | 0.64 (0.18) |

$q_{PHB}$ in Cmol PHB/Cmol $X$ d; $-q_S$ in Cmol Acet/Cmol $X$ d; $-q_{PO4}$ in Pmol $PO_4$/Cmol $X$ d; $Y_{PHB/S}$ in Cmol PHB/Cmol Acet; $Y_{X/S}$ in Cmol $X$/Cmol Acet
[a] Average values calculated from 2 SBR cycles
[b] Average values calculated from 5 SBR cycles

Indeed, under low light SBR operation, the culture continued to favour growth instead of polymer accumulation as indicated by the higher growth yield on acetate ($Y_{X/S}$ of 0.52 ± 0.06 Cmol $X$/Cmol Acet) in comparison to the PHA production yield per acetate consumed (only $Y_{PHB/S}$ of 0.03 ± 0.02 Cmol PHB/Cmol Acet). Interestingly, the culture maintained its substrate and polymer kinetic performance despite the lower volumetric illumination of the selector SBR. Perhaps this can be explained by the internal illumination of the present reactor that may have led to a more homogenous light distribution, thus being more efficiently supplied to the biomass (in [11] the reactor was externally illuminated. See section 2.1). Regardless, the culture was enriched in purple phototrophic bacteria (Supplementary material Figure S1, S2) that could accumulate PHA during accumulation trials. Precisely to determine the best conditions to operate the accumulator reactors, several operating parameters were sequentially tested, starting with the effect of the initial carbon concentration on the culture performance.

## 3.2 Effect of initial substrate concentration

In some mixed microbial culture PHA producing processes, the culture selection step can be followed by an accumulation step, in a separate reactor, to further increase the PHA content of the biomass. In the present study, where the selector SBR is operated under a permanent carbon feast regime, the mixed liquor that is withdrawn at the end of the cycle and directed to the accumulation step, always contains residual carbon. To evaluate the effect of the initial carbon concentration on the culture performance during the accumulation step, two acetate concentrations were tested. One corresponded to the concentration typically present in the SBR at the beginning of the reactor cycle after feeding (Test A - higher concentration of $70 \pm 0.5$ CmM). The other corresponded to the concentration left by the residual acetate present in the mixed liquor withdrawn from the SBR at the end of the cycle (Test B - lower concentration of $28 \pm 0.6$ CmM). Results indicate that the culture presents higher acetate uptake rate and PHB production rate when operated with a lower initial acetate concentration (Figure 1).

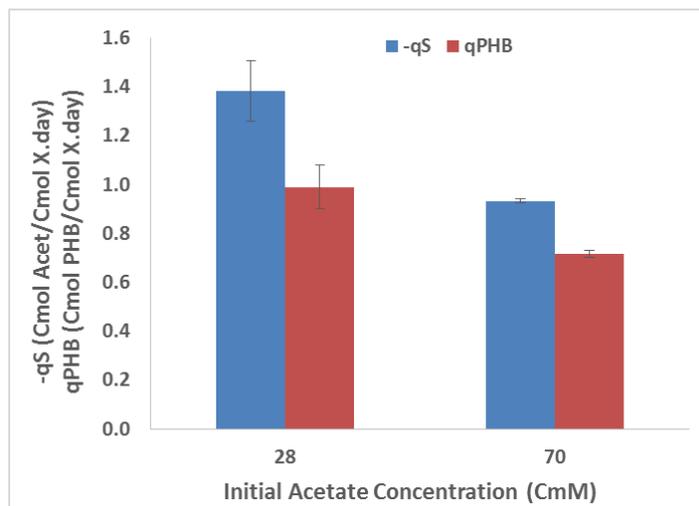

Fig 1 – Specific acetate uptake rate and specific PHB production rate in relation to the acetate concentration at the beginning of the accumulation tests. Error bars calculated from duplicate tests.

In fact, the acetate uptake rate increased 48% from 0.93 ± 0.01 to 1.38 ± 0.12 Cmol Acet/Cmol $X$ d when the initial acetate concentration was decreased by 60%. As for the PHB production rate, it increased 38% from 0.72 ± 0.01 to 0.99 ± 0.09 Cmol PHB/Cmol $X$ d. Although the increase of the acetate uptake rate was higher than the increase of the PHA production rate (leading to a slight decrease in the PHA production yield from 0.77 ± 0.02 to 0.72 ± 0.00 Cmol PHB/Cmol Acet), the system improved its overall polymer productivity.

A possible explanation for this culture response to lower acetate concentrations may be that higher acetate concentrations can be inhibitory to the culture, decreasing its metabolic rates. This result points to the benefit of operating accumulator reactors with a low concentration of carbon source. Also it points to a possible advantage of changing the presently used dump feed SBR operation to a continuous mode in order to operate the culture with lower substrate concentrations. Regarding the biomass growth (monitored by ammonia consumption along the tests), no noticeable growth rate differences were observed between tests with the culture, presenting ammonia consumption rates of 0.02

± 0.01 Nmol NH$_4$/Cmol $X$ d and 0.03 ± 0.01 Nmol NH$_4$/Cmol $X$ d, at higher and lower acetate concentrations, respectively.

With the results from this test, it was decided that further accumulation trials should be conducted with low acetate concentration.

### 3.3 Effect of light availability

In phototrophic processes, light is known to be one of the major factors impacting the culture's performance. In previous works with PHA accumulating PMCs, accumulation tests were conducted with light intensities that varied amongst works, but a consistent and specific evaluation of the effect of the light intensity on the culture metabolism was never performed. For this light intensity study, the accumulator reactors were operated in the same manner as in Test B of the previous section, 3.2, i.e., starting with low acetate concentrations, around 28 CmM. Four increasing volumetric light intensities were tested up to a volumetric light intensity of 20.7 ± 0.7 W/ L. This highest value tested approaches the maximum values that can be naturally obtained and was chosen based on the typical light intensities that can be found with sunlight illumination. In some areas of the globe, sun can illuminate the surface of the Earth with an irradiance that can reach instantaneous values of 900 - 1000 W/m$^2$ but for example, in the case of the southern Iberian Peninsula, values up to 700 – 800 W/m$^2$ can be observed for 3 to 4 hours during Autumn and up to 8 hours during Summer ([15,16], Supplementary material Figure S1). Due to the low penetration depth of light and in order to maximize light distribution, phototrophic reactors can have small depths (e.g. 4 – 10 cm) that lead to high illuminated surface/volume ratios of 10 to 25 m$^2$/m$^3$ (assuming, for example, a flat panel reactor design). This combined with the southern Iberian Peninsula irradiance means that there can be maximum operating volumetric light intensities from 7 to 20 W/L, depending on the time of the day and on the reactor depth. With this in mind, four volumetric light intensities were tested which resulted in the specific light intensities of 0.96 ± 0.04 W/g *X*, 4.1 ± 0.04 W/g *X*, 10.2 ± 0.4 W/g *X* and 18.5 ± 1.3 W/g *X* (Figure 2).

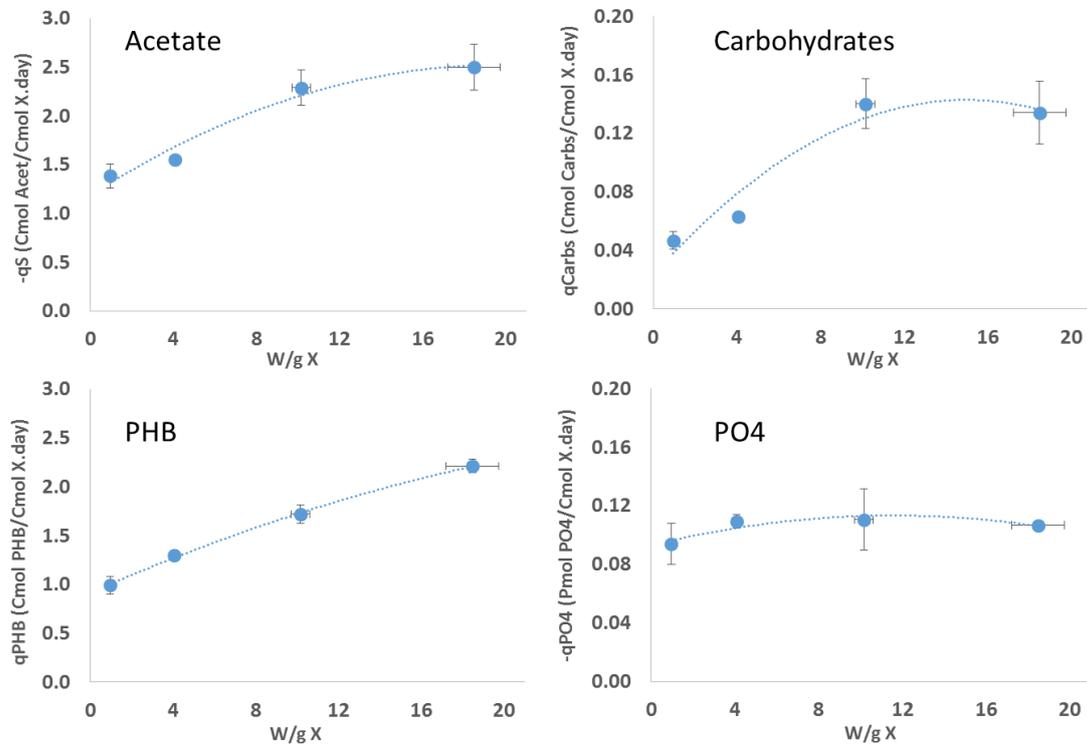

Figure 2 – Impact of specific light intensity on the specific acetate and phosphate uptake rates and in the specific PHB and carbohydrates production rates. Error bars calculated from duplicate tests. The average initial acetate and ammonia concentration was 28.1 ± 0.9 CmM and 11.7 ± 0.3 NmM, respectively, for the eight tests.

Results indicate that the specific acetate uptake rate increased along with the light intensity but decelerated when reaching the highest specific light intensity tested of 18.5 ± 1.3 W/g *X*. It could be that the membrane acetate transporters were becoming saturated [17] or the phototrophic systems were approaching their maximum electron transfer capacity at this point, not being able to provide more ATP for acetate uptake [18,19]. In any case, the maximum value obtained for the specific acetate uptake rate was 2.50 ± 0.23 Cmol Acetate/Cmol *X* d, which is an 80% increase in relation to the lowest light intensity tested. Also, it is the maximum value ever reported for PMCs operated under permanent carbon feast strategies since in previous work, values of 1.5 ± 0.3 Cmol Acetate/Cmol *X* d were attained at specific light intensities around 6 W/g *X* [11].

Regarding the polymer production, the specific PHB production rate increased with the light intensity very much in line with the specific acetate uptake rate. A similar deceleration is noticed at the highest light intensity, likely linked to the acetate uptake deceleration. This culture behaviour towards higher light intensities was also observed in pure cultures of purple bacteria (*Rhodobacter sphaeroides*) producing hydrogen [20,21]. In this case, the hydrogen production rate also increased with the light availability up to a point where further increase in the light intensity did not change the production rate. In the present work, the maximum PHB production rate was obtained at the highest light intensity tested, which resulted in a specific PHB production rate of $2.21 \pm 0.07$ Cmol PHB/Cmol $X$ d. Again, this is the highest value ever reported for PMCs, being three times higher than the value observed in the previously mentioned work ($0.73 \pm 0.13$ Cmol PHB/Cmol $X$ d). Also, it narrows the difference with the specific PHB production rates observed in accumulation tests with aerobic organisms selected with acetate. While in previous works PMCs presented PHB production rates ten to twenty times lower than the ones obtained with aerobic organisms, in this study a less than threefold difference was observed. Indeed, [22] obtained a rate of 6 Cmol PHB/Cmol $X$ d with acetate enriched aerobic organisms in accumulation tests with comparable conditions (initial acetate concentration of 30 CmM and non-growth limiting ammonia concentration of 2.8 NmM). These are very promising results for PMCs.

Regarding the PHB storage yield, results indicate that an average of $0.83 \pm 0.07$ Cmol PHB/Cmol Acet can be obtained with PMCs at the highest light intensities (Figure 3).

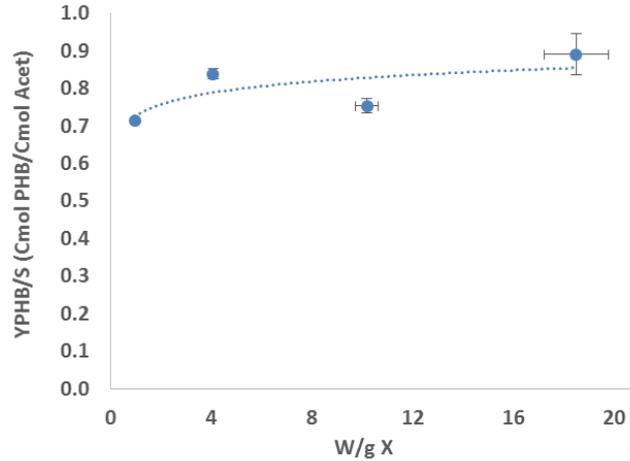

Figure 3 - PHB storage yield on acetate obtained for the tests at different light intensities.

These values are higher than the ones observed in aerobic systems. Aerobic organisms, typically present PHB production yields on acetate around 0.4 – 0.6 Cmol PHB/Cmol Acet [22,23] since their ATP production is dependent on carbon oxidation which leads to carbon loss by decarboxylation. In contrast, PMCs are operated anaerobically and are composed of purple phototrophic bacteria that can obtain ATP from light, thus enabling a high conversion efficiency of acetate into PHB while minimizing $CO_2$ production. This conversion efficiency combined with higher kinetic rates at higher light intensities, enabled, in the present work, PHB content increases from 15% to 30% PHB/VSS in less than 4 hours, which fits within the timeframe for maximum sunlight intensity exposure (Figure S3). These results are an indication that natural sunlight illumination can fulfil the energetic demand of PMCs necessary for PHA production.

The light intensity studies also provided some insight regarding carbohydrates production. While the culture tripled its carbohydrates production, this rate reached a plateau even before the maximum light intensity was tested, with values around 0.14 ± 0.02 Cmol Carbs/Cmol $X$ d at a light intensity of 10.2 ± 0.4 W/g $X$ and carbohydrate production yields throughout the light intensity study of just 0.05 ± 0.01 Cmol

Carbs/Cmol Acet. This is an indication that the culture truly prefers to store carbon as PHB and that there is a low risk of the culture shifting its storage metabolism to carbohydrates when higher light intensities are provided during the accumulation step.

Finally, although results indicate that the culture responded to the increase of light intensity with an overall increase of its metabolic rates, the specific P uptake rate was constant along the tests. It should be noted, however, that phosphate is typically removed in the first 6 to 7 hours of the selector SBR cycle operation (Supplementary material Figure S4) with the PMC presenting high yields of P removal per biomass growth ($Y_{P/X}$ ~ 0.2 to 0.4), indicating possible poly-phosphate accumulation (as also observed in [11]). Indeed, if the taken up P was solely used for cell growth it would be expected an $Y_{P/X}$ of only 0.02 Pmol $PO_4$/Cmol $X$. Therefore, Figure 2 results suggest that the culture is removing phosphate at its maximum capacity immediately from the time zero of the tests, and therefore, its phosphate removal rate is independent of the light availability.

The results from the effect of light intensity on the culture performance indicate a direct correlation between the culture metabolism and light availability, with the highest PHB production capacity being attained at the higher light intensity of 18.5 ± 1.3 W/g $X$. Thus, it was settled that further accumulation tests should be conducted at specific light intensities around 20 W/g $X$.

**3.4 Effect of residual nitrogen removal**

The third condition that was tested in this work was the removal of the residual ammonia that accompanies the mixed liquor withdrawn from the selector SBR. It was hypothesised that by removing the N availability, the culture would not be capable of growing during the accumulation step and therefore, the taken up carbon could be solely used for PHB accumulation. In combination with a high light availability (20 W/L) this could lead to a further increase of the PHB production rate and storage yield in accumulation assays.

Results indicated that when the mixed liquor was centrifuged, the supernatant was removed and the obtained biomass was re-suspended in fresh medium (Table 3, condition a)), an overall decrease of the metabolic rates were observed in comparison with the previous tests (section 3.3, 20 W/L), where the withdrawn mixed liquor was not subjected to this pre-treatment (Table 3, condition b)).

Table 3 - Kinetic parameters of accumulating tests with different initial mixed liquor and feeding conditions. Mixed liquor and medium fractions are in relation to the total reactor operating volume. $[Acet]_i$ in condition b) and c) corresponds to the residual acetate in the mixed liquor.

| Conditions | | $[Acet]_i$ | $X_i$ | W/L | W/g $X$ | $-q_S$ | $q_{PHB}$ | $q_{Carbs}$ | $-q_{PO4}$ |
|---|---|---|---|---|---|---|---|---|---|
| a) Supernatant removed + full make up with medium lacking acetate and NH4 | | pulses, < 3.5 CmM | 41 | 20 | 19.8 | 1.16[a] (0.14) | 0.61 | 0.07 | 0.03 |
| b) 2/3 mixed liquor + 1/3 medium lacking acetate | | 27 (0.4) | 46 (1.6) | 20.7 (0.7) | 18.5 (1.3) | 2.50 (0.23) | 2.21 (0.07) | 0.13 (0.02) | 0.11 (0.00) |
| c) Only mixed liquor | First part: consumption of residual acetate | 41 | 68 | 20 | 12.0 | 2.40 | 1.40 | 0.12 | - |
| | Second part: acetate pulse addition | pulses, < 3.5 CmM | 72 | 20 | 11.3 | 2.43[b] (0.48) | 1.40 | 0.04 | - |

$[Acet]_i$ - initial concentration; $X_i$ in Cmmol/L; $q_{PHB}$ in Cmol PHB/Cmol $X$ d; $-q_S$ in Cmol Acet/Cmol $X$ d; $-q_{PO4}$ in Pmol $PO_4$/Cmol $X$ d; $q_{Carbs}$ in Cmol Carbs/Cmol $X$ d; Values in condition b are the average values of two tests with standard deviation in brackets.

[a] Average of 6 acetate pulse additions

[b] Average of 10 acetate pulse additions

Despite both tests being conducted under a very similar specific light intensity (~19 W/g *X*), all specific uptake and production rates were negatively affected under condition a), with this condition particularly impacting the PHB production rate and on the final PHB content that did not go further than 20% PHA/VSS. It could be the case that cell centrifugation at 9000 g, supernatant removal (removal of signalling molecules) or the complete volume make up with fresh medium (higher ratio of phosphate, minerals or metals per microorganism than usually observed in the SBR) could have inhibited the culture. Thus, it was decided to evaluate the culture performance in conditions where the biomass in the mixed liquor had not been affected by centrifugation, supernatant removal or fresh medium addition (Table 3, condition c)). It should be noted however, that by not adding fresh medium, biomass was not diluted, leading to higher initial biomass concentration, and consequently, lower specific light intensity. Results indicate a slightly lower acetate uptake rate in relation to condition b) during the consumption of residual acetate, which agrees with Figure 2 that shows a slightly lower rate at specific light intensities around 12 W/g *X*. Also, the PHB production rate agrees with Figure 2, indicating a decrease of PHB production rate with the specific light intensity decrease. These results suggest that using the mixed liquor directly from the SBR also leads to metabolic rates that seem to correlate with the specific light intensity applied. As such, there seems to be no negative effect of diluting the mixed liquor with up to 1/3 of fresh medium, as conducted in previous sections.

Furthermore, to rule out the possible influence of acetate pulse addition on the condition a) culture performance, the culture under condition c) was also pulse fed after the residual acetate had been completely consumed. Overall there seems to be no significant difference in the culture kinetic parameters when being pulse fed at low acetate concentrations or when continuously consuming the residual acetate.

Although it could not be evaluated the culture performance in the absence of $NH_4$ *per se*, these tests reveal that mixed liquor over-dilution or centrifugations with supernatant removal will lead to decreased metabolic rates. No inhibition is observed when the mixed liquor is diluted with up to 1/3 of fresh medium or when pulse-fed with acetate concentrations < 3.5 CmM.

**3.5 PHA accumulation with fermented cheese whey as substrate**

In the previous sections, the PMC was solely fed with acetate, the substrate to which the culture had been acclimatized. This resulted in the production of a PHA homopolymer containing only HB monomers. PHB polymers are brittle but their properties can be improved by incorporating HV monomers which creates a HB:HV co-polymer with lower melting point and reduced brittleness [2,24]. In order to evaluate the culture capability of using a real fermented waste (cheese whey) as substrate and simultaneously producing heteropolymeric PHA, it was decided to conduct two accumulation tests, one with fermented cheese whey (FCW) and the other with synthetic FCW. The synthetic FCW was used with the intention of evaluating if unknown components present in the real waste could impact on the culture behaviour (composition of FCW is depicted in Table 1). Results indicated the culture preference for acetate consumption, being this substrate completely consumed at each pulse, while the other compounds tended to accumulate (Table 4).

Table 4 – Average uptake rates of the different compounds during accumulation tests calculated for the six pulses of synthetic FCW or FCW.

|  | Lactic Acid | Acetic Acid | Propionic Acid | Butyric Acid | Valeric Acid | Ethanol | Total |
|---|---|---|---|---|---|---|---|
| Synthetic FCW | 0.05 (0.02) | 0.83 (0.12) | 0.07 (0.02) | 0.05 (0.04) | 0.01 (0.01) | 0.02 (0.01) | 1.04 (0.18) |
| FCW | 0.01 (0.01) | 0.85 (0.05) | 0.06 (0.01) | 0.03 (0.02) | 0.00 (0.00) | 0.03 (0.02) | 0.98 (0.05) |

Tests were conducted with an illumination of 20 W/L and $X_i$ = 43 CmM that led to a specific illumination of 18.9 W/g $X$. Rates in Cmol/Cmol $X$ d and the standard deviation for each uptake rate is shown in brackets.

This is an anticipated result since the PMC had been enriched with acetate. Nevertheless, and despite the fact that the culture had never been acclimated to other carbon sources, the biomass took up the other compounds immediately from the first pulse addition, with the exception of FCW valeric acid that was never consumed by the culture and only accumulated.

Overall, Table 4 shows no significant difference in feeding the culture with synthetic or real FCW. The total specific substrate uptake rate was on average 1.0 Cmol substrate/Cmol $X$ d for both tests, with acetate contributing for more than 80% of this rate. Despite the similar substrate uptake rates, the test with synthetic FCW led to a PHA production rate of 0.60 Cmol PHA/Cmol $X$ d (15% of HV monomers production) while with FCW this was only 0.53 Cmol PHA/Cmol $X$ d (12% of HV monomers) (Figure 4).

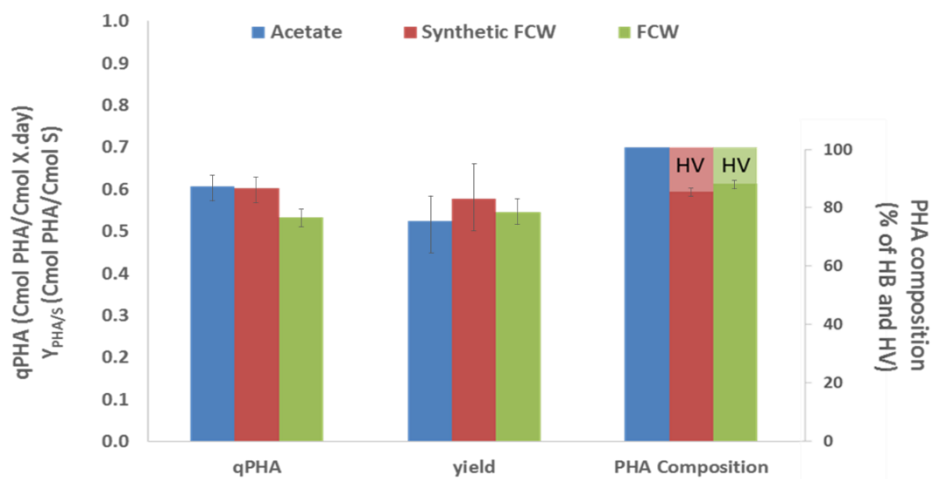

Figure 4 – Comparison between specific PHA production rate, PHA storage yield and PHA composition in accumulation tests with pulses of acetate, synthetic FCW and FCW. In the PHA composition bars, lighter and darker colours indicate, respectively, HV and HB percentage in the produced polymer.

This resulted in a PHA content variation from 10 to 25% in the test with synthetic FCW and from 10 to 20% in the FCW test, for the same period of time. This lower PHA accumulation with real FCW could be related to the influence of other compounds present in the FCW that eventually diverted the taken up carbon from the PHA metabolic pathway or enabled the growth of non-PHA storing organisms. In any case, for both tests the culture showed its capability of producing a PHA co-polymer containing HB and HV monomers. The HV content ranged between 12% - 15% (% Cmol) which is within the range of HV content of 5% - 30% obtained in accumulation tests with aerobic mixed cultures selected with FCW [12,25]. This is a good indication of the potential of PMCs to produce PHA co-polymers as long as a suitable substrate containing HV precursors (like propionate) is provided.

When the tests with synthetic and FCW are compared with the test with acetate pulses (Table 3, condition a)) we see that the specific PHB production rate with synthetic FCW is very similar to the test with acetate pulses (0.61 Cmol PHB/Cmol $X$ d) (Figure 4). However, in this last case, and due to the substrate nature, it was only obtained an HB homopolymer that also varied from 10 to 25% in content. Regarding the PHA storage yield, no significant differences were observed when the culture was either fed with synthetic FCW, FCW or acetate (values in the range of 0.55 - 0.60 Cmol PHA/Cmol substrate). The same occurred with the final PHA content that ranged around 20%-25% PHA/VSS for the three substrates. When comparing these results with studies that used aerobic mixed microbial cultures (MMCs) fed with FCW, we see that MMCs presented higher storage yields (0.7 – 0.8 Cmol PHA/Cmol substrate) and PHA contents between 65% - 75% PHA/VSS [25, 26, 27]. However, these studies were conducted with cultures that had been selected with FCW. In the present work, the PMC was selected with acetate and had not been previously acclimatized to the acids/ethanol present in the FCW. Yet, when the culture was fed with FCW, it was capable of attaining PHA storage yields around 0.6 Cmol PHA/Cmol substrate, showing that PMCs are robust, adapting well to new substrates and with storage yields within the ranges observed with MMCs fed with real substrates [5,8]. More importantly, PMCs can accomplish this with no aeration requirement which can lead to the potential decrease in PHA production costs.

Overall these tests extended the PMC concept to real substrates and results show that an acetate enriched PMC was capable of accumulating a PHA co-polymer using real FCW with specific PHA production rates and final PHA content comparable to synthetic FCW and acetate feeding. Future work will focus on selecting the culture directly with a real fermented waste to acclimatize the organisms to other compounds and follow the feed-shift impact on the microbial population and metabolic performance. This will allow the comparison of PMCs PHA productivity and economic feasibility in relation to other technological approaches.

**4 Conclusion**

The development of PMC systems for PHA production has led to a recent operational proposal of sequentially selecting the culture under low light demanding conditions, followed by an accumulation step under high light intensity. This work showed the feasibility of this operating strategy with results indicating that by acting on the operating conditions of accumulator reactors it was possible to increase PHA production efficiency. Low acetate concentrations and high operating light intensities in accumulator reactors led to specific PHA production rates of $2.21 \pm 0.07$ Cmol PHB/Cmol $X$ d, which enabled content increases from 15 to 30% in less than 4h. Also, it was shown for the first time the culture capability of using a real waste, fermented cheese whey, to produce PHA with an HV content of 12%. These results prospect the utilization of VFA-rich fermented wastes as substrates for PHA production with PMCs and open up the possibility for direct sunlight illumination, since the light intensities used in this work (20 W/L) can be naturally obtained in sunny regions.

**Supplementary material of this work can be found in online version of the paper**

**Supplementary Material**

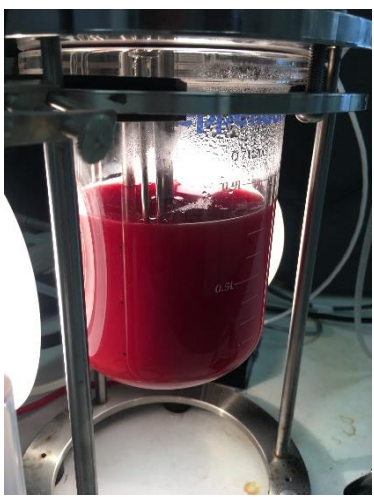

Figure S1 – Picture of the PMC showing the purple colour of phototrophic purple bacteria.

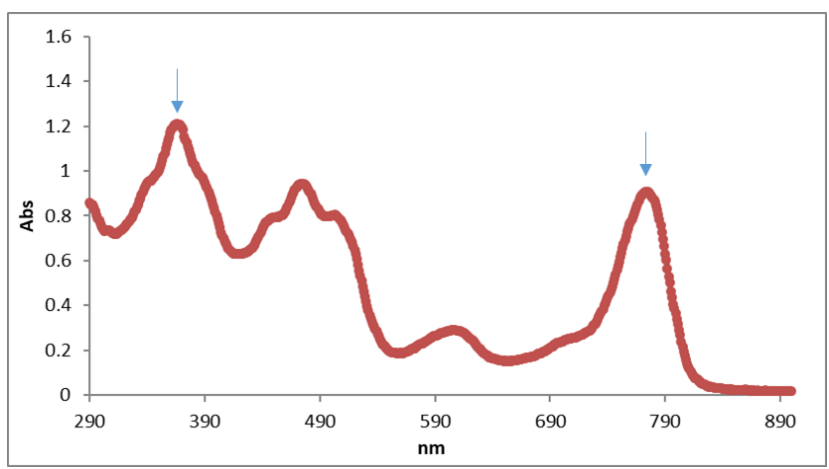

Figure S2 – Ultraviolet, visible and infrared light absorbance spectrum of the pigments extracted from the PMC in ethanol. Absorbance peaks of bacteriochlorophyll (typical pigment of purple bacteria) are shown at 366 nm and 775 nm.

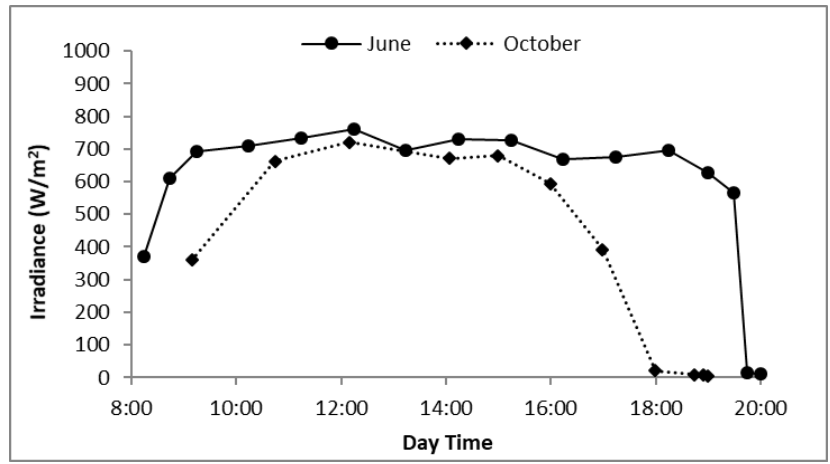

Figure S3 – Solar irradiance variation along the day time. Data was registered on a 30º tilted plane turned south on the 13-October-2014 and 20-June-2016 in Caparica, Portugal (38°39'46.7"N 9°12'29.1"W).

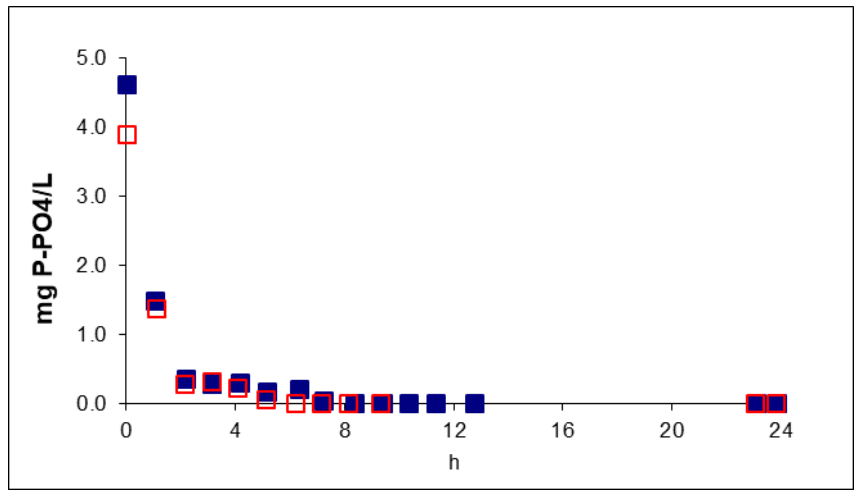

Figure S4 – Phosphate consumption profile typically observed in the SBR. Two cycles of SBR are represented by the two markers, blue square and open square.

**Acknowledgments**

The authors wish to acknowledge Ms Virgínia Carvalho for assistance with solar irradiance measurements. The authors would also like to acknowledge the Fundação para a Ciência e Tecnologia (Portugal) for funding through SFRH/BPD/101642/2014. Applied Molecular Biosciences Unit - UCIBIO acknowledges financing by national funds from FCT/MEC (UID/Multi/04378/2013) and co-financed by ERDF under PT2020 Partnership Agreement (POCI-01-0145-FEDER-007728). NoAW project received funding from the European Research Council (ERC) under the European Union's Horizon 2020 research and innovation programme (grant agreement n° 688338).

# References

[1] Amass, W., Amass, A., Tighe, B., 1998. A Review of Biodegradable Polymers: Uses, Current Developments in the Synthesis and Characterization of Biodegradable Polyesters, Blends of Biodegradable Polymers and Recent Advances in Biodegradation Studies. Polym. Int. 47, 89–144.

[2] Arcos-Hernández, M. V., Laycock, B., Donose, B.C., Pratt, S., Halley, P., Al-Luaibi, S., Werker, A., Lant, P., 2013. Physicochemical and mechanical properties of mixed culture polyhydroxyalkanoate (PHBV). Eur. Polym. J. 49, 904–913.

[3] Chandra, R., Rustgi, R., 1998. Biodegradable polymers. Prog. Polym. Sci. 23, 1273–1335.

[4] Rodriguez-Perez, S., Serrano, A., Pantión, A.A., Alonso-Fariñas, B., 2018. Challenges of scaling-up PHA production from waste streams. J. Environ. Manage. 205, 215–230.

[5] Kourmentza, C., Plácido, J., Venetsaneas, N., Burniol-Figols, A., Varrone, C., Gavala, H.N., Reis, M.A.M., 2017. Recent Advances and Challenges towards Sustainable Polyhydroxyalkanoate (PHA) Production. Bioengineering 4, 55.

[6] Reis, M.A.M., Albuquerque, M., Villano, M., Majone, M., 2011. Mixed Culture Processes for Polyhydroxyalkanoate Production from Agro-Industrial Surplus/Wastes as Feedstocks, in: Moo-Young, M., Fava, F., Agathos, S. (Eds.), Comprehensive Biotechnology. Academic Press, Burlington, pp. 669–683.

[7] Reis, M.A.M., Serafim, L.S., Lemos, P.C., Ramos, A.M., Aguiar, F.R., Van Loosdrecht, M.C.M., 2003. Production of polyhydroxyalkanoates by mixed microbial cultures. Bioprocess Biosyst. Eng. 25, 377–85.

[8] Nikodinovic-Runic, J., Guzik, M., Kenny, S.T., Babu, R., Werker, A., O'Connor, K.E., 2013. Carbon-rich wastes as feedstocks for biodegradable polymer (polyhydroxyalkanoate) production using bacteria, Chapter4, 1st ed, Advances in Applied Microbiology. Vol 84, Elsevier Inc.

[9] Fradinho, J.C., Domingos, J.M.B., Carvalho, G., Oehmen, A., Reis, M. A. M., 2013. Polyhydroxyalkanoates production by a mixed photosynthetic consortium of bacteria and algae. Bioresour. Technol. 132, 146–153.

[10] Fradinho, J., Oehmen, A., Reis, M., 2013. Effect of dark/light periods on the polyhydroxyalkanoate production of a photosynthetic mixed culture. Bioresour. Technol. 148, 474–479.

[11] Fradinho, J.C., Reis, M.A.M., Oehmen, A., 2016. Beyond feast and famine: Selecting a PHA accumulating photosynthetic mixed culture in a permanent feast regime. Water Res. 105, 421–428.

[12] Gouveia, A.R., Freitas, E.B., Galinha, C.F., Carvalho, G., Duque, A.F., Reis, M.A.M., 2017. Dynamic change of pH in acidogenic fermentation of cheese whey towards polyhydroxyalkanoates production: Impact on performance and microbial population. N. Biotechnol. 37, 108–116.

[13] Lanham, A., Ricardo, A., Coma, M., Fradinho, J., Carvalheira, M., Oehmen, A., Carvalho, G., Reis, M.A.M., 2012. Optimisation of glycogen quantification in mixed microbial cultures. Bioresour. Technol. 118, 518–525.

[14] APHA, AWWA, WPCF, 1995. Standard Methods for the Examination of Water and Wastewater. American Public Health Association, Washington DC.

[15] Gschwind, B., Ménard, L., Albuisson, M., Wald, L., 2006. Converting a successful research project into a sustainable service: The case of the SoDa Web service. Environ. Model. Softw. 21, 1555–1561. http://www.soda-is.com.

[16] Sancho, J.M., Riesco, J., Jiménez, C., Sánchez de Cos, M.C., Montero, J., López, M., 2012. Atlas de Radiación Solar en España utilizando datos del SAF de Clima de EUMETSAT. Agencia Estatal de Meteorología (AEMET), Ministerio de Agricultura, Alimentación y Medio Ambiente del Gobierno de España. http://www.aemet.es/es/serviciosclimaticos/datosclimatologicos/atlas_radiacion_solar

[17] Imhoff J.F. (1995) Taxonomy and Physiology of Phototrophic Purple Bacteria and Green Sulfur Bacteria. In: Blankenship R.E., Madigan M.T., Bauer C.E. (eds) Anoxygenic Photosynthetic Bacteria. Advances in Photosynthesis and Respiration, vol 2. Springer, Dordrecht

[18] Polle, J.E.W., Kanakagiri, S., Jin, E.S., Masuda, T., Melis, A., 2002. Truncated chlorophyll antenna size of the photosystems - A practical method to improve microalgal productivity and hydrogen production in mass culture. Int. J. Hydrogen Energy 27, 1257–1264.

[19] Timpmann, K., Chenchiliyan, M., Jalviste, E., Timney, J.A., Hunter, C.N., Freiberg, A., 2014. Efficiency of light harvesting in a photosynthetic bacterium adapted to different levels of light. Biochim. Biophys. Acta - Bioenerg. 1837, 1835–1846.

[20] Nath, K., Das, D., 2009. Effect of light intensity and initial pH during hydrogen production by an integrated dark and photofermentation process. Int. J. Hydrogen Energy 34, 7497–7501.

[21] Uyar, B., Eroglu, I., Yucel, M., Gunduz, U., Turker, L., 2007. Effect of light intensity, wavelength and illumination protocol on hydrogen production in photobioreactors. Int. J. Hydrogen Energy 32, 4670–4677.

[22] Serafim, L.S., Lemos, P.C., Oliveira, R., Reis, M.A.M., 2004. Optimization of polyhydroxybutyrate production by mixed cultures submitted to aerobic dynamic feeding conditions. Biotechnol. Bioeng. 87, 145–60.

[23] Jiang, Y., Hebly, M., Kleerebezem, R., Muyzer, G., Loosdrecht, M.C.M. Van, 2011. Metabolic modeling of mixed substrate uptake for polyhydroxyalkanoate (PHA) production. Water Res. 45, 1309–1321.

[24] Ten, E., Jiang, L., Zhang, J., Wolcott, M.P., 2015. Mechanical performance of polyhydroxyalkanoate (PHA)-based biocomposites. In Biocomposites: Design and Mechanical Performance, Pages 39-52. Woodhead Publishing

[25] Oliveira, C.S.S., Silva, C.E., Carvalho, G., Reis, M.A., 2017. Strategies for efficiently selecting PHA producing mixed microbial cultures using complex feedstocks: Feast and famine regime and uncoupled carbon and nitrogen availabilities. N. Biotechnol. 37, 69–79.

[26] Duque, A.F., Oliveira, C.S.S., Carmo, I.T.D., Gouveia, A.R., Pardelha, F., Ramos, A.M., Reis, M.A.M., 2014. Response of a three-stage process for PHA production by mixed microbial cultures to feedstock shift: Impact on polymer composition. N. Biotechnol. 31, 276–288.

[27] Colombo, B., Sciarria, T.P., Reis, M., Scaglia, B., Adani, F., 2016. Polyhydroxyalkanoates (PHAs) production from fermented cheese whey by using a mixed microbial culture. Bioresour. Technol. 218, 692–699.